\documentclass{PoS}
\usepackage{booktabs}
\usepackage{amsmath}
\usepackage{marginnote}
\usepackage{simplewick}
\usepackage{tikz}
\usepackage{graphics}
\usepackage{gincltex}
\usepackage{bm}
\usepackage[backend=biber,url=false,doi=false,giveninits=true,hyperref,style=numeric,sorting=none,maxnames=1]{biblatex}
\renewbibmacro{in:}{}
\AtEveryBibitem{\clearfield{title}}
\usetikzlibrary{decorations.markings}
\usetikzlibrary{shapes.multipart}
\usetikzlibrary{shapes.geometric}
\usetikzlibrary{svg.path}
\usepackage{pgfplots}
\pgfplotsset{compat=1.6}

\newcommand{\eval}[1]{\langle#1\rangle}
\newcommand{\Tr}{\operatorname{Tr}}
\newcommand{\tr}[1]{\operatorname{Tr}\{#1\}}
\newcommand{\bigO}[1]{\mathrm{O}(#1)}
\newcommand{\Nf}{N_\mathrm{f}}

\newcommand{\Doo}{D_\mathrm{oo}}
\newcommand{\Dee}{D_\mathrm{ee}}
\newcommand{\Deo}{D_\mathrm{eo}}
\newcommand{\Doe}{D_\mathrm{oe}}

\title{Multi-level integration for meson propagators}

\ShortTitle{Multi-level integration for meson propagators}

\author{Leonardo~Giusti$^a$, \speaker{Tim~Harris}$^a$, \speaker{Alessandro~Nada}$^b$, Stefan~Schaefer$^b$\\
        \llap{$^a$}Dipartimento di Fisica, Universit\`a di Milano-Bicocca, and INFN, Sezione di Milano-Bicocca\\
        Piazza della Scienza 3, I-20126 Milano, Italy\\
        \llap{$^b$}John von Neumann Institute for Computing, DESY\\
        Platanenallee 6, D-15738 Zeuthen, Germany\\
        E-mail: \email{leonardo.giusti@mib.infn.it}, \email{tharris@mib.infn.it},\email{alessandro.nada@desy.de}, \email{stefan.schaefer@desy.de}}

\abstract{
The computation of many correlation functions in lattice QCD is severely
hindered by a signal-to-noise problem.
Recent developments in the factorization of both the fermion propagator and
determinant pave the way for the implementation of multi-level Monte Carlo
integration techniques for lattice QCD.
In these proceedings we introduce new strategies for the estimation of the
factorized contribution to the connected and disconnected diagrams for meson
two-point functions.
An estimator for the factorized connected diagram is constructed sequentially
for a two-level integration scheme.
For the disconnected diagram, we introduce an improved estimator by performing
a frequency-splitting of traces, applicable with or without multi-level
integration.
Preliminary results in the quenched theory with a two-level integration scheme
are presented.
\vskip1cm
DESY 18-186
}

\FullConference{The 36th Annual International Symposium on Lattice Field Theory - LATTICE2018\\
		22-28 July, 2018\\
		Michigan State University, East Lansing, Michigan, USA.}

\begin{document}

\section{Introduction}
\label{sec:introduction}

In the context of lattice Quantum Chromodynamics (QCD), numerical computations
of correlation functions suffer from the fact that the signal
decreases with the distance exponentially faster that the statistical error~\cite{Parisi:1983ae,Lepage:1989hd}.
This well-known signal-to-noise ratio problem currently limits the precision
of several determinations, such as the hadronic vacuum polarization
contribution (HVP) to the muon $g-2$, masses and matrix elements of baryon
states, amplitudes of leptonic and semi-leptonic $B$ meson decays and many
others.

A possible solution to this problem can be found in the framework of
multi-level integration techniques: if the action and the observable can be
factorized, independent measurements on different domains can be combined
together, so that the noise decreases with the distance exponentially faster
with respect to a standard Monte~Carlo simulation.
In the context of pure-gauge theories, this approach has proved to work
extremely
well~\cite{Parisi:1983hm,Luscher:2001up,Meyer:2002cd,DellaMorte:2007zz,DellaMorte:2008jd,DellaMorte:2010yp},
essentially either taming or completely solving the problem.

However, the non-locality of both the determinant and the quark propagator makes an extension to fermionic theories highly non-trivial. 
In the last few years a conceptual advancement in this
direction~\cite{Ce:2016idq,Ce:2016ajy,Giusti:2017ksp,Ce:2017ndt} provided the
theoretical framework for an implementation of multi-level techniques in full
QCD.
In this contribution we focus on recent developments in this line of research
for the computation of the observables in a multi-level fashion by introducing
novel techniques to calculate both connected and disconnected two-point
functions.

For what concerns the connected diagrams, we exploit the possibility of
placing random sources on a time-slice in between the source and the sink 
so to compute the two-point function using a sequential propagator.
In this approach the average over the first region can be easily taken, thus
paving the way for a two-level simulation with two domains.

In the case of disconnected diagrams, we introduce a new variance-reduction
strategy for the computation of disconnected quark loops, which can also be
applied to ordinary QCD simulations.
Then we investigate the disconnected diagram for the meson two-point function
in a two-level simulation, and confirm that a simple factorization
scheme~\cite{Ce:2016idq} is applicable also in the case of the two-point
function of the vector current.

These proceedings are structured as follows: in section~\ref{sec:snr_ml} we
analyse the signal-to-noise ratio problem and its possible solution in the
form of multi-level simulations, while in section~\ref{sec:factorization} we
introduce the factorization of the quark propagator first proposed
in~\cite{Ce:2016idq}.
Sections~\ref{sec:connected} and \ref{sec:disconnected} represent the main
contribution of these proceedings, in which we present the latest developments
concerning the computation of connected and disconnected diagrams in a
multi-level simulation.
Finally, in section~\ref{sec:conclusions} we draw some conclusions.

\section{Signal-to-noise ratios and multi-level integration}
\label{sec:snr_ml}

A typical example is a meson correlation function at non-zero momentum $\bf p$, which at large enough source-sink separation $\lvert y_0 - x_0\rvert$ is 
\begin{align}
\label{eq:snr_1}
C_\Gamma (x_0,y_0,{\bf p}) \propto \exp \left[ - E_\Gamma ({\bf p}) \lvert y_0 - x_0\rvert \right] 
\end{align}
while the variance of this object decays with the lightest state with four quark lines, e.g. the two-pion state
\begin{align}
\sigma^2_\Gamma (x_0,y_0,{\bf p}) \propto \exp \left[ -2 M_\pi \lvert y_0 - x_0\rvert \right] 
\end{align}
leading to an exponentially vanishing signal-to-noise ratio,
\begin{align}
\label{eq:snr_3}
 \frac{C_\Gamma (x_0,y_0,{\bf p})}{\sigma_\Gamma (x_0,y_0,{\bf p})/\sqrt{n}} \propto \sqrt{n} \, \exp \left[ - (E_\Gamma ({\bf p}) - M_\pi) \lvert y_0 - x_0\rvert \right],
\end{align}
where $n$ is the number of gauge configurations.

In a multi-level Monte~Carlo simulation the observable is sampled in two (or more) levels: at level-0, $n_0$ configurations are generated as in a standard computation; at level-1, for each level-0 configuration $n_1$ configurations are generated by updating independently different space-time regions using the factorized action.
Eventually, the factorized estimator of the observable, such as a two-point function, for a two-level simulation over $m$ domains can be written as
\begin{align}
 C^{\text{(f)}}_\Gamma (x_0,y_0,{\bf p}) \simeq \frac{1}{n_0} \sum_{i=1}^{n_0} \left[O_1\right]_{n_1} \left[O_2\right]_{n_1} ... \left[O_m\right]_{n_1}
\end{align}
where each term of the type $\left[O_k\right]_{n_1}$ represents the inner level-1 average over the $k$-th space-time region.
Ideally, the standard error of this estimator would then scale like $n_0^{-1/2} n_1^{-m/2}$, thus enhancing \textit{exponentially} the signal-to-noise ratio of eq.~\ref{eq:snr_3}. 

\section{Factorization of quark propagator}
\label{sec:factorization}

We are interested in the factorization of the quark propagator $Q^{-1} (y,x)$, a non-local functional of the gauge field over the entire lattice.
Following~\cite{Ce:2016idq,Giusti:2017ksp}, we start by introducing in the temporal direction three non-overlapping domains $\Lambda_i$ ($i=1,2,3$) that encompass the full lattice; in this particular decomposition the domain $\Lambda_0 \cup \Lambda_2$ is a disconnected one (i.e. $\Lambda_0$ and $\Lambda_2$ do not touch).
In the following we will refer to $\Lambda_1$ as the ``frozen'' region, while to $\Lambda_0$ and $\Lambda_2$ as the ``active'' regions.

In this geometrical setup the Hermitian Wilson-Dirac operator can be written in block form as
\begin{align}
 Q=
\begin{pmatrix}
 Q_{\Lambda_{0,0}} & Q_{\Lambda_{0,1}} & 0 \\
 Q_{\Lambda_{1,0}} & Q_{\Lambda_{1,1}} & Q_{\Lambda_{1,2}} \\
 0 & Q_{\Lambda_{2,1}} & Q_{\Lambda_{2,2}} \\
\end{pmatrix}
.
\end{align}
A possible, simple decomposition is defined by taking the overlapping domains $\Omega_i^*= \Lambda_i \cup \Lambda_{i+1}$, so that in each of them the $Q$ operator becomes
\begin{align}
 Q_{\Omega_i^*}=
\begin{pmatrix}
 Q_{\Lambda_{i,i}} & Q_{\Lambda_{i,i+1}} \\
 Q_{\Lambda_{i+1,i}} & Q_{\Lambda_{i+1,i+1}} \\
\end{pmatrix}
.
\end{align}
It was shown in~\cite{Ce:2016idq,Giusti:2017ksp} that the exact quark propagator from a source $x \in \Lambda_0$ to a sink $y \in \Lambda_2$ can be written as
\begin{align}
\label{eq:exact_qp_1}
 Q^{-1} = - Q_{\Omega_1^*}^{-1} Q_{\Lambda_{1,0}} Q^{-1}
\end{align}
or, equivalently
\begin{align}
\label{eq:exact_qp_2}
 Q^{-1} = - Q_{\Omega_1^*}^{-1} Q_{\Lambda_{1,0}} \frac{1}{1-\omega} Q_{\Omega_0^*}^{-1}
\end{align}
where the operator $\omega$ is defined as $\omega=Q_{\Omega_0^*}^{-1} Q_{\Lambda_{1,2}} Q_{\Omega_1^*}^{-1} Q_{\Lambda_{1,0}}$.
It is crucial to observe that the dependence on the gauge fields of the two overlapping domains is made explicit.
By taking the first term of the geometric series of eq.~\ref{eq:exact_qp_2}, one can write an approximate, fully-factorized version of the propagator as
\begin{align}
\label{eq:fact_qp}
 Q^{-1} \simeq - Q_{\Omega_1^*}^{-1}  Q_{ \Lambda_{1,0}}  Q_{\Omega_0^*}^{-1}.
\end{align}

On the other hand, in the case of $x,y \in \Lambda_0$ one can write the quark propagator as
\begin{equation}
 \label{eq:exact_qp_disc}
 Q^{-1} = Q_{\Omega_0^*}^{-1} + Q_{\Omega_0^*}^{-1} Q_{\Lambda_{1,2}} Q_{\Omega_1^*}^{-1} Q_{\Lambda_{1,0}} \frac{1}{1-\omega} Q_{\Omega_0^*}^{-1}
\end{equation}
which is instrumental for the computation of disconnected diagrams.

\section{Connected diagrams}
\label{sec:connected}

We study the correlation function $C_\Gamma$ of two non-diagonal quark densities
\begin{align}
\label{eq:two_pt_fct}
 C_\Gamma(y_0,x_0) =  \frac{a^6}{L^3} \sum_{\mathbf{x},\mathbf{y}} \eval{\overline d(y) \Gamma u(y) \, \overline u(x) \Gamma d(x) }
                   = -\frac{a^6}{L^3} \sum_{\mathbf{x},\mathbf{y}} \langle W_\Gamma (y,x) \rangle
\end{align}
where $W_\Gamma (y,x)$ is the Wick contraction
\begin{align}
\label{eq:exact_mp}
 W_\Gamma (y,x) = \Tr \left\{ Q^{-1} (y,x) \Gamma \left[ Q^{-1} (y,x) \right]^\dagger \Gamma \right\}.
\end{align}

In this section we present the definition of a new estimator of the meson correlator with stochastic sources: indeed, if we rewrite the quark propagator $Q^{-1}$ as in eq.~\eqref{eq:exact_qp_1} we obtain
\begin{align}
\label{eq:exact_mp_2}
 W_\Gamma (y,x) = \Tr \left\{ Q_{ \Lambda_{1,0}} \, Q^{-1} (\cdot,x) \, \Gamma \, Q^{-1} (x, \cdot) \, Q_{ \Lambda_{0,1}} \, Q_{\Omega_1^*}^{-1} (\cdot,y) \, \Gamma \, Q_{\Omega_1^*}^{-1}(y,\cdot)  \right\}.
\end{align}
Upon closer inspection of eq.~\eqref{eq:exact_mp_2}, it is clear that the standard choice of positioning the random vectors at the coordinate $x_0$, i.e. the time-slice on which the source is located (as in eq.~\eqref{eq:two_pt_fct}), is not the only one. 
In fact, one could place them on the interior boundary of the $\Lambda_0$ domain, i.e. exactly on the time slice next to the second domain $\Lambda_1$, and thus introduce a new estimator for the two-point function.

Furthermore, if we use the \textit{approximate} quark propagator of eq.~\eqref{eq:fact_qp} we can introduce a fully factorized meson correlation function whose contraction is
\begin{align}
\label{eq:fact_wick}
 W^{\text{(f)}}_\Gamma (y,x) = \Tr \left\{  \left[ Q_{ \Lambda_{1,0}} \, Q_{\Omega_0^*}^{-1} (\cdot,x) \, \Gamma \, Q_{\Omega_0^*}^{-1} (x, \cdot) \, Q_{ \Lambda_{0,1}} \right] \left[ \, Q_{\Omega_1^*}^{-1} (\cdot,y) \, \Gamma \, Q_{\Omega_1^*}^{-1}(y,\cdot) \right] \right\}
\end{align}
where one can easily note that the term $Q_{\Omega_0^*}^{-1}(\cdot,x) \,
\Gamma \, Q_{\Omega_0^*}^{-1}(x,\cdot)$ involves only fields belonging to
$\Omega_0^*$, while $Q_{\Omega_1^*}^{-1}(\cdot,y) \, \Gamma \,
Q_{\Omega_1^*}^{-1}(y,\cdot)$ only fields belonging to $\Omega_1^*$.
We stress the fact that the factorized two-point function
$C^{\text{(f)}}_\Gamma$ obtained from eq.~\eqref{eq:fact_wick} is an
approximation: an unbiased estimator for the exact correlator requires the
computation of the \textit{rest}, defined as
\begin{align}
 C^{\text{(r)}}_\Gamma(y_0,x_0) = C_\Gamma(y_0,x_0) - C^{\text{(f)}}_\Gamma(y_0,x_0).
\end{align}

In practice, we can use the freedom in positioning the noise source to define a new estimator $ \tilde W^{\text{(f)}}_\Gamma$ for the Wick contraction of the factorized two-point function, namely
\begin{align}
\label{eq:fact_wick_ns}
 \nonumber
 \tilde W^{\text{(f)}}_\Gamma (y,x) = \frac{1}{N_\eta}
 \sum_{i=1}^{N_\eta} & \sum_{\substack{\bm z,\bm w,\bm{w'}\\\bm{z''},\bm{z'}}}
  \eta_i(z)^\dagger \, Q_{ \Lambda_{0,1}}(z,w) \, Q_{\Omega_1^*}^{-1}(w,y) \, \Gamma \,  Q_{\Omega_1^*}^{-1}(y,w') \\
    & \times Q_{ \Lambda_{1,0}}(w',z'') \, Q_{\Omega_0^*}^{-1}(z'',x) \, \Gamma \, Q_{\Omega_0^*}^{-1}(x, z') \, \eta_i(z') 
\end{align}
where all the space-time coordinates have been made explicit: $z_0 = z'_0 =
z''_0$ is the time coordinate of the interior boundary of $\Lambda_0$, where
the random vectors are located, while $w_0 = w'_0$ is the interior boundary of $\Lambda_1$.
As usual, the random fields $\eta_i(x)$ ($i=1,...,N_{\eta}$) are defined such
that
\begin{align}
  \label{eq:noise}
  \langle \eta_i (x)\rangle_{\eta} &= 0,\\
  \langle \eta_i (x) \eta_j (y)^\dagger \rangle_{\eta} &=
    \delta_{ij} \delta_{xy}
\end{align}
where spin and colour indices have been suppressed.

The Wick contraction of eq.~\eqref{eq:fact_wick_ns} can be computed in a
Monte~Carlo multi-level integration scheme by evaluating the meson propagator sequentially.
The strategy of the calculation requires that the propagator matrix element (spin and colour indices are suppressed)
\begin{align}
    \phi_i (w')= Q_{ \Lambda_{1,0}}(w',\cdot) \, Q_{\Omega_0^*}^{-1}(\cdot,x)
    \, \Gamma \, Q_{\Omega_0^*}^{-1}(x,\cdot) \, \eta_i (\cdot)
\end{align}
is computed first for each of the $n_1$ level-1 configurations of the $\Omega_0^*$ domain; the level-1 average $\eval{\phi_i}_{n_1}$ is stored for each noise vector.
Then, for each of the level-1 configurations of the $\Omega_1^*$ region one computes
\begin{align}
    Q_{\Omega_1^*}^{-1}(y,\cdot) \, Q_{ \Lambda_{1,0}}(\cdot,\cdot)
    \,\eta_i(\cdot)  \qquad \text{ and } \qquad Q_{\Omega_1^*}^{-1}(y,\cdot) \,
    \left[\phi_i(\cdot)\right]_{n_1} 
\end{align}
where $\left[\ldots\right]_{n_1}$ indicates the level-1 average previously computed on $\Omega_0^*$. 
From these two quantities the correlation function is computed at the sink $y_0$ averaging over the $n_1$ level-1 configurations.

\subsection{Numerical tests for the vector correlator}
\label{sub:vector_ml}

In order to assess the feasibility of this technique and its effectiveness in dealing with the signal-to-noise ratio problem, a numerical study is performed in the quenched theory.
In particular, we compute the factorized vector correlator at zero momentum (averaged over $\gamma_i$, $i=1,2,3$) both in a conventional Monte~Carlo simulation, to assess the size and the error of the rest of the propagator, and in a two-level simulation, to test the effective gain in the simulation cost.

\begin{table}[t]
  \centering
  \begin{tabular}{cccccccc}
  \toprule
  \rule{0pt}{9pt} chain id & $\beta$ & $L/a$ & $T/a$ & $\kappa$ &
  $c_\mathrm{sw}$ & $m_\mathrm{PS}$\,(MeV) & $a$\,(fm) \\
  \midrule
  \rule{0pt}{9pt} D2 & $6.2$ & 32 & 96 &  $0.1352$ & $1.61375$ & $580$ & $0.068$ \\
  \bottomrule
  \end{tabular}
 \caption{\label{tab:num_setup} Numerical setup of Monte~Carlo simulation test in the quenched approximation.} 
\end{table}
Using the numerical setup reported in tab.~\ref{tab:num_setup}, $n_0=100$
level-0 configurations are generated with in quenched QCD with the Wilson
action and periodic boundary conditions in all directions.
For the generation of the level-1 configurations, the lattice is divided in the time direction into the two active regions
\begin{itemize}
 \item $\Lambda_0 = \{ x : x_0 \in (0 ,39)a \}$,
 \item $\Lambda_2 = \{ x : x_0 \in (48,87)a \}$;
\end{itemize}
and into the frozen region 
\begin{itemize}
 \item $\Lambda_1 = \{ x : x_0 \in (40,47)a \} \cup \{ x : x_0 \in (88,95)a \}$,
\end{itemize}
which is composed of two disconnected domains, each of width $\Delta=8$.
For each level-0 configuration and for each active region belonging to $\Omega_0^*$ and $\Omega_1^*$, $n_1=16$ level-1 configurations were produced while keeping the fields in $\Lambda_1$ fixed.
The connected two-point function is computed using $N_\eta=40$ noise sources
placed in the interior boundaries of $\Lambda_0$, namely $z_0=0$ and $z_0=39a$.
The factorized contribution, $C^{\text{(f)}}$, is calculated using
eq.~\eqref{eq:fact_wick_ns} both with and without multi-level integration. 
In order to compute the rest, $C^{\text{(r)}}$, the exact two-point function is obtained using the same noise sources for the exact propagator of eq.~\eqref{eq:exact_qp_1}. 

\begin{figure}[t]
  \centerline{
    \includegraphics[scale=.8]{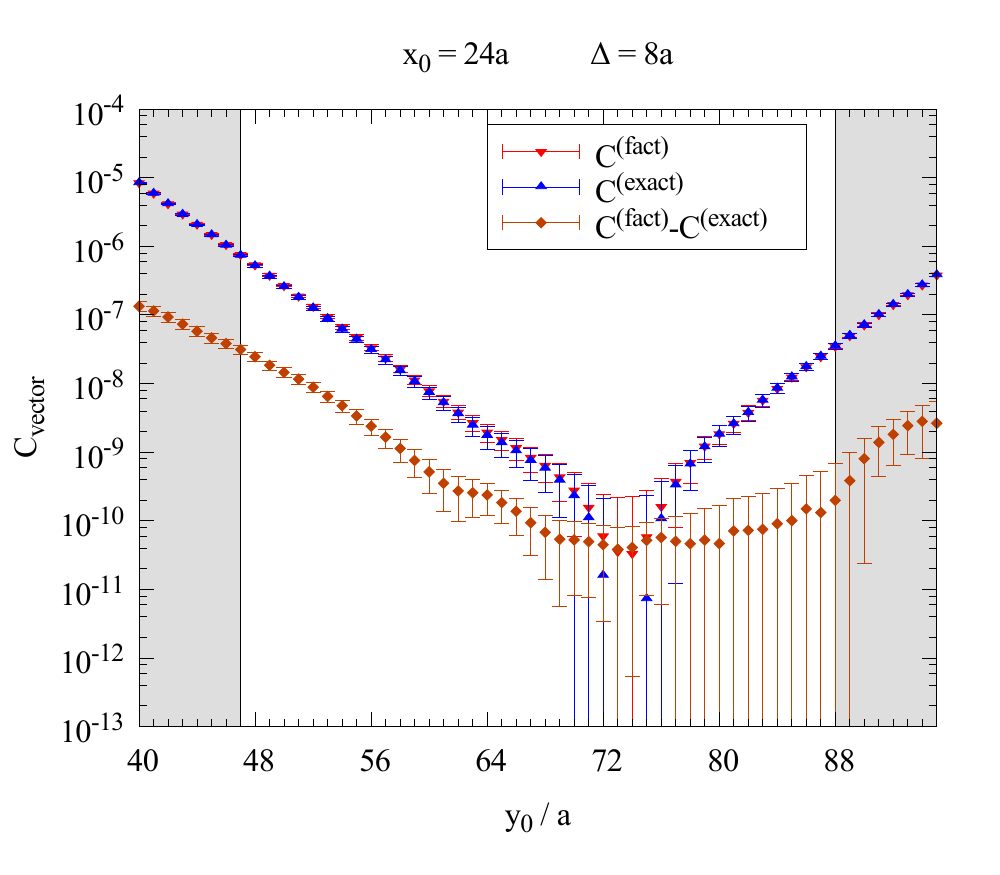}%
    \includegraphics[scale=.8]{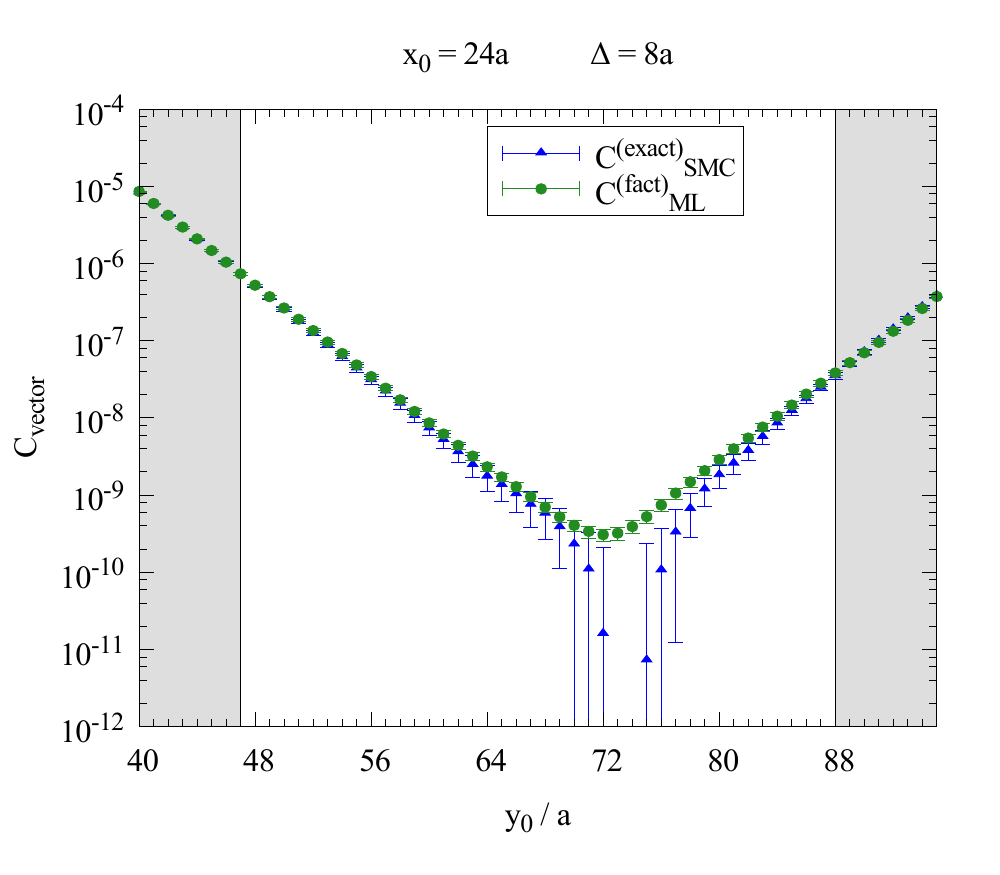}
  }
  \caption{Left: results for the exact vector correlator (red triangles), along with the factorized contribution (blue triangles) and the rest (dark-red triangles). 
          Right: comparison of results for the exact vector correlator
        calculated with a standard Monte~Carlo simulation (SMC, blue
      triangles), and the factorized contribution computed with a multi-level integration scheme (ML, green dots).}
  \label{fig:vector_ml}
\end{figure}
In the left panel of fig.~\ref{fig:vector_ml} we present results obtained with $x_0=24a$ for the exact
two-point function, its factorized part, computed in a standard Monte~Carlo
(SMC), and the rest of the propagator. 
We first observe the appearance of the signal-to-noise ratio problem at source-sink separations $\lvert y_0 - x_0 \rvert $ of around 40 lattice spacings. 
Furthermore, the size of the rest of the propagator is relatively small, namely about 5\% of the value of the exact correlation function. 
As long as the signal is present, the exact and factorized results agree within error. 
We remark that these results are obtained with all $n_0 \times n_1$
configurations: since the level-1 configurations share the same frozen
regions, the data analysis must be performed by binning the $n_1$ configurations.
Then, we proceed with the extraction of results within a multi-level integration scheme: from the right panel of fig.~\ref{fig:vector_ml} it is immediately clear how this approach is able to deal with the signal-to-noise ratio problem, drastically reducing the error while using the same $n_0\times n_1$ configurations.

\begin{figure}[t]
  \centerline{
    \includegraphics[scale=.8]{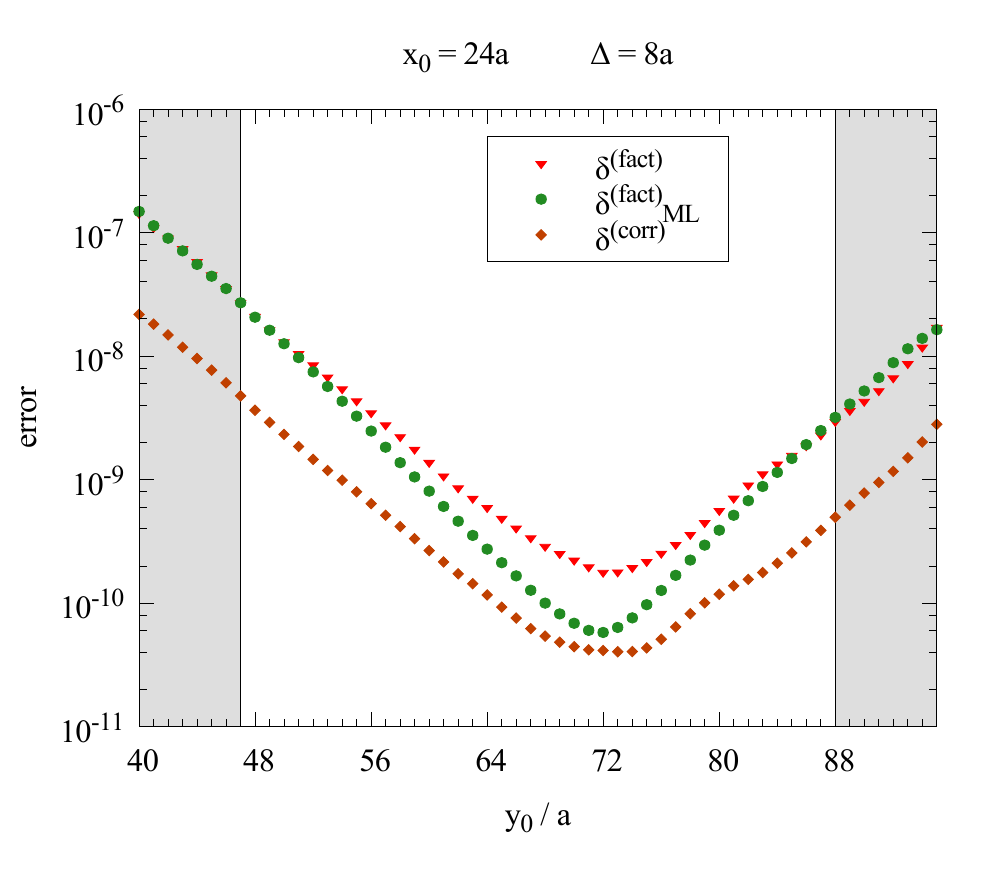}%
    \includegraphics[scale=.8]{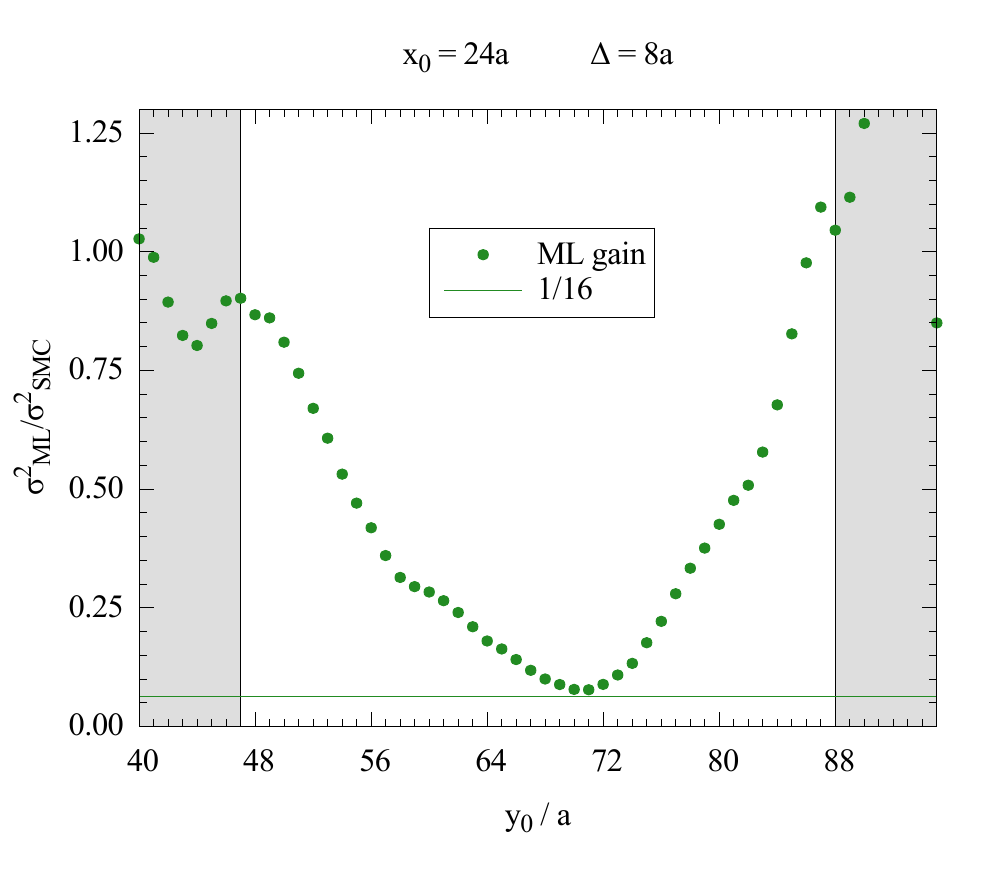}
  }
  \caption{Left: comparison of errors associated to the factorized vector
    correlator with (green dots) and without (red triangles) multi-level, along with the error associated to the rest (dark-red triangles).
           Right: ratio between the variance obtained with a multi-level scheme and with a standard Monte~Carlo simulation, on the same set of $n_0 \times n_1$ configurations.}
  \label{fig:vector_error}
\end{figure}
The left panel of fig.~\ref{fig:vector_error} clearly shows how the error is
reduced by making use of the multi-level technique described in section~\ref{sec:connected}. 
Furthermore, we can observe that, in this numerical setup, the error on the
rest of the two-point function is still smaller that the error on the
factorized part, even when obtained in a multi-level scheme with $n_1=16$.
Finally, in order to understand what is the gain in terms of computational
cost, one can look at the ratio of the variance obtained with a standard
Monte~Carlo and with a multi-level technique. 
The ``ideal'' scaling one expects would be $1/n_1$: in the right panel of
fig.~\ref{fig:vector_error} we can observe that this ratio depends strongly on
the distance from the frozen region, and that indeed it reaches the expected
scaling of $1/n_1=1/16$ when far away enough from $\Lambda_1$.

\section{Disconnected diagrams}
\label{sec:disconnected}

Quark-line disconnected diagrams arise in any correlation function with fermion
bilinears which contain some flavour-singlet component, such as the
electromagnetic current, whose matrix elements appear in the computation of
hadronic form factors, and the leading hadronic contributions to muon $g-2$.
Furthermore, disconnected diagrams are also present in meson spectroscopy in
isosinglet channels, and in generalized susceptibilities at finite temperature
and density.
Several of these observables suffer from the signal-to-noise problems described
earlier and it is crucial to test multi-level methods for these correlation
functions.

In the following subsection we propose a variance reduction method for the
stochastic estimation of disconnected quark loops and test it
numerically in a single-level quenched simulation with the same parameters
used in sec.~\ref{sub:vector_ml}.
In subsection~\ref{sub:disc_multi} we investigate the disconnected
contribution to the vector two-point function in a two-level simulation using
the proposed variance-reduction methods.

\subsection{Variance-reduction for disconnected diagrams}
\label{sub:freq}

The disconnected quark loop is usually computed with a noisy estimator for the
trace~\cite{doi:10.1080/03610919008812866},
\begin{align}
  \tr{\Gamma S^f(x,x)} &\approx
  \begin{cases}
    \mathrm{Im}{T^f_{\Gamma}(x,x)} \quad
    & \textrm{if }\Gamma=\gamma_\mu,\\
      \mathrm{Re}{T^f_{\Gamma}(x,x)} \quad
      & \textrm{if }\Gamma\in\{1,\gamma_5,\gamma_\mu\gamma_5,\sigma_{\mu\nu}\},
  \end{cases}\\
  T^f_{\Gamma}(x,x) &= \frac{1}{N_\eta}\sum_{n=1}^{N_\eta}
      \eta_n^\dagger(x) \Gamma \sum_y S^f(x,y)\eta_n(y),
  \label{eq:hutchinson}
\end{align}
where $S=Q^{-1}\gamma_5$ is the $\gamma_5$-Hermitian quark propagator,
$\Gamma$ is a Dirac matrix, and the independent components of the noise
$\eta_n(x)$ are drawn from a $\mathrm{U}(1)$ distribution, which has zero mean
and finite variance, as per eqs.~\eqref{eq:noise}.
Although this estimator is unbiased, it converges only with
$\bigO{N_\eta^{-1/2}}$ to the expectation value, namely that its variance
scales inversely proportional to the number of sources.
However, in order to test the scaling of the gauge noise in a two-level
integration scheme we ought to have a precise estimate of the quark loop.

\begin{figure}[t]
  \centerline{
    \includegraphics{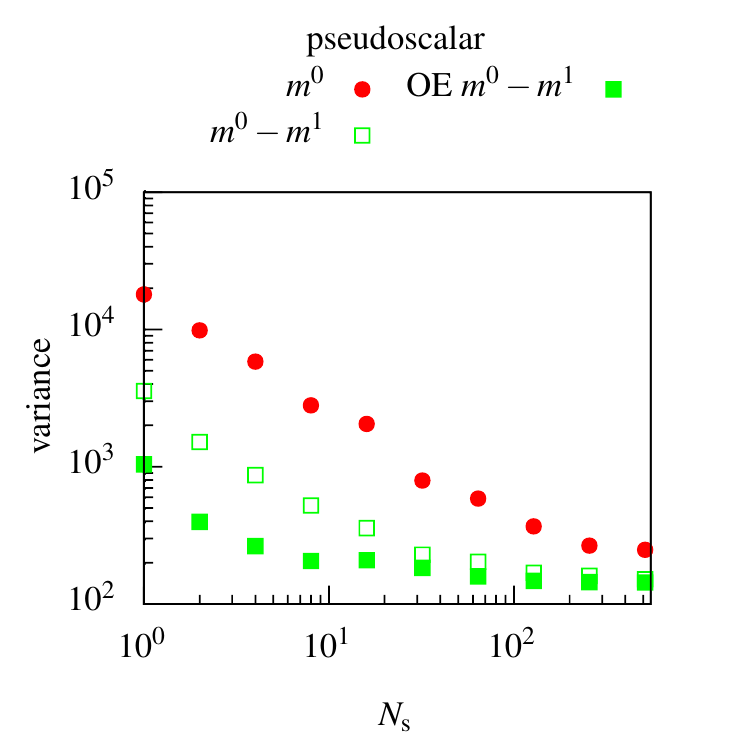}%
    \includegraphics{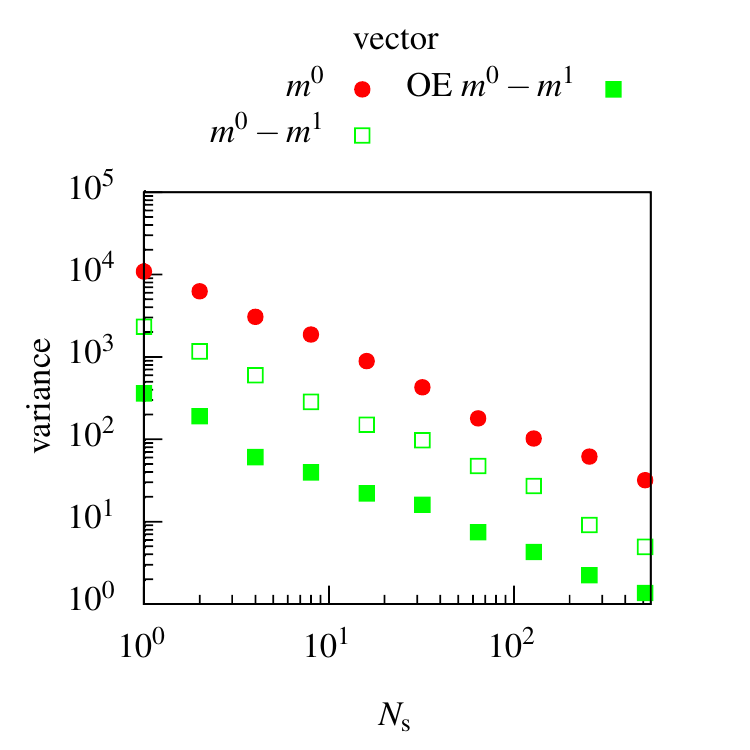}
  }
  \caption{The variance as a function of the number of sources,
  $N_\eta$, in the pseudoscalar (left) and vector (right) channels. The
red symbols represent the variance for the light quark mass while the green
symbols the difference between the light and a heavier quark mass.
The open and filled green symbols are two different estimators for the
difference, explained in the text.}
  \label{fig:disc_oneend}
\end{figure}
As can be seen in fig.~\ref{fig:disc_oneend}, the variance of the quark loop
summed on a timeslice (red circles) scales inversely to the number of sources.
In the pseudoscalar channel (left), the variance begins to saturate to the
gauge noise after a few hundred noise sources, while in the vector channel
(right) no evidence for the reaching the gauge limit is observed.
This compels us to search for an improved estimator, especially for the vector
channel, for which the gauge noise has, to the best of our knowledge, never
been reached.

An alternative estimator can be constructed for the single-quark loop by
performing a frequency-splitting of the trace, inspired by the
mass-preconditioning of the forces in the HMC~\cite{Hasenbusch:2002ai}, by
constructing the telescoping sum
\begin{align}
  \tr{\Gamma S^{f_0}(x,x)} &= \tr{\Gamma (S^{f_0}-S^{f_1})(x,x)} + \tr{\Gamma
  (S^{f_1}-S^{f_2})(x,x)}
  + \ldots + \tr{\Gamma S^{f_M}(x,x)},
  \label{eq:freq}
\end{align}
with a hierarchy of bare quark masses $m^{f_0}_0<m^{f_1}_0<\ldots<m^{f_M}_0$.
If each of the traces on the right-hand size of the equation is computed
separately via
\begin{align}
  \label{eq:ls_diff}
  \tr{\Gamma (S^{f_i}-S^{f_{i+1}})(x,x)}
  &\approx \mathrm{Re}(\mathrm{Im})\left[
    \frac{1}{N_\eta}\sum_{n=1}^{N_\eta}
  \eta_n^\dagger(x) \Gamma
  \sum_y \{S^{f_i}(x,y)-S^{f_{i+1}}(x,y)\}\eta_n(y)
  \right],
\end{align}
then the frequency-splitting estimator effectively splits the sources of the
variance from different energy scales.
If the dominant contribution is from the UV part, then the noise can be
controlled at larger quark masses which requires less computational effort,
much like integrating the computationally intensive forces on a larger
time-scale in the HMC.

Furthermore, we expect a large correlation between propagators with similar
quark masses which means that the estimate eq.~\eqref{eq:ls_diff} will have a
reduced variance to the single quark loop~\cite{Meyer:2018til}.
Indeed, the variance of the estimator for the difference is reduced with
respect to the variance on a single trace, as can be observed in
figure~\ref{fig:disc_oneend} (green open squares), where a second flavour with
a large quark mass $m^1_0$ has been introduced, see tab.~\ref{tab:ym_masses}.

Another variation can be obtained by using an identity for the Wilson-Dirac
operator to express the difference of a doublet of propagators with different
quark masses as a product
\begin{align}
  S^{f_i}(x,y) - S^{f_{i+1}}(x,y) &=
(m^{f_{i+1}}_0 - m^{f_i}_0) \sum_z S^{f_{i+1}}(x,z)S^{f_i}(z,y).
   \label{eq:oneend}
\end{align}
%
The product on the right-hand side gives a new possibility for an estimator
for difference eq.~\eqref{eq:ls_diff} via
\begin{align}
  \nonumber
  \tr{\Gamma (S^{f_{i}}-S^{f_{i+1}})(x,x)}
  &\approx (m^{f_{i+1}}_0 - m^{f_{i}}_0)\times\\
  &\quad\mathrm{Re}(\mathrm{Im})\left[\frac{1}{N_\eta}\sum_{n=1}^{N_\eta}
  \tr{\Gamma
\sum_{y,z} S^{f_{i+1}}(x,y)\eta_n(z)\eta^\dagger_n(z)S^{f_{i}}(z,x)}\right].
    \label{eq:oneend_diff}
\end{align}
While this one-end estimator (OE) has the same expectation value as
eq.~\eqref{eq:ls_diff}, it is clear that they are not equal sample-by-sample,
and indeed they have different variances%
\footnote{See ref.~\cite{Boucaud:2008xu} for a discussion of the variance of a
  one-end estimator for the connected diagram.}.
This can be verified by a short computation of the variance in terms of the
correlation functions after integrating out the noise fields.
In practice, we observe a further reduction in both the pseudoscalar and
vector channels using this estimator, as seen in fig.~\ref{fig:disc_oneend}
(green filled squares).
Using CLS $\Nf=2$ configurations, we have observed a large reduction in the
variance when the quark masses are set to the light valence and strange quark
masses, and therefore the application of this identity to compute the
disconnected contribution to the HVP is very promising~\cite{Giusti:2019inprep}.

In the following we propose a frequency-splitting estimator (FS) using the
one-end estimator with $M=2$ auxiliary masses, with the largest bare subtracted
mass on the order of the lattice cutoff scale, see tab.~\ref{tab:ym_masses}.
We employ the $k=4$ order hopping parameter expansion to evaluate the term
$\tr{\Gamma S^{f_2}(x,x)}$ via
\begin{align}
S^f(x,y) &= \sum_{n=0}^{k-1}H^n(\Dee + \Doo)^{-1}
        + (H^kS^f)(x,y)
\end{align}
where $H=-(\Doo^{-1}\Doe + \Dee^{-1}\Deo)$ is the hopping matrix.
We use either spatial dilution or hierarchical probing
vectors~\cite{Stathopoulos2013} for the first `hopping' terms, and a standard
Monte Carlo for the remainder.
The advantage of this set-up is that the stochastic noise is completely
eliminated in the estimation of the hopping terms after
$k^4/8$ quadratures, if $k/2$ is a power of two.
The use of the hopping expansion in the mass-preconditioned HMC has also
recently been explored in ref.~\cite{Hasenbusch:2018iuw}.

\begin{figure}[t]
  \centerline{
    \includegraphics{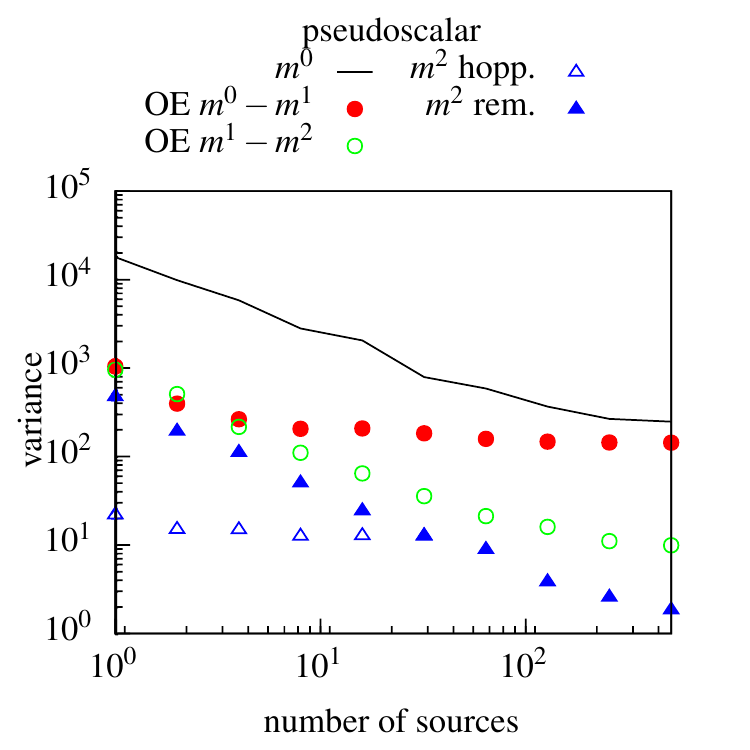}%
    \includegraphics{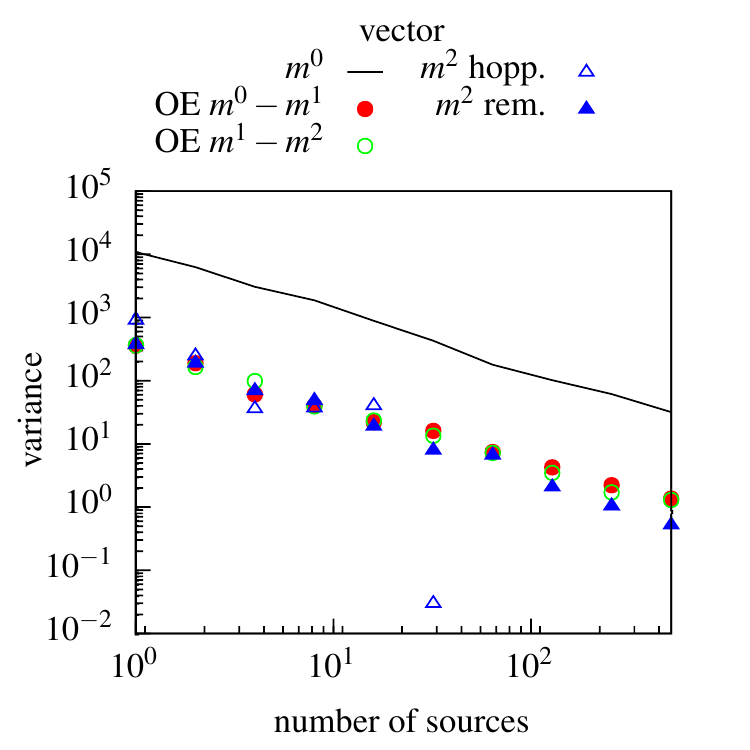}
  }
  \caption{The variance versus the number of sources for each of the
  contributions to the frequency-splitting (FS) estimator (points) and the
standard Monte Carlo (solid line) in the pseudoscalar (left) and vector
(right) channels. Note that the hopping terms, which do not require inversions
of the Dirac matrix, are computed using hierarchical probing with spin-colour
dilution.}
  \label{fig:disc_splittings}
\end{figure}
The variance of all of the components of this estimator are shown in
fig.~\ref{fig:disc_splittings} as a function of the number of noise sources,
along with the same for the standard stochastic estimator
eq.~\eqref{eq:ls_diff} for the light quark loop (solid line).
In the pseudoscalar channel, the variance on the one-end estimator for the
light doublet dominates the variance (filled red circles), while in the vector channel, the
hopping term is initially dominant (open blue triangles).
Note that the hopping terms are estimated exactly at the right-most open
triangle with $k^4/8=32$ quadratures.

Without any particular fine-tuning, it is a straightforward and cheap
procedure to measure the variances on the each component of the FS estimator
with just a few sources.
In this way the cost can be optimized for a given observable.
The inversions used in the estimator of the smallest quark-mass difference can
be reused to create the standard estimator and compare the cost of the methods
a posteriori.
Furthermore, we expect that this estimator should behave well toward the
chiral limit, as it effectively performs a separation of the sources of the
variance at the IR and UV scales, which appears to dominate in the vector
channel.
In fact, as the light quark inversions become more expensive, the cost of the
exact computation of the hopping terms is amortized, being independent on the
quark mass.

\subsection{Numerical results in the two-level integration}
\label{sub:disc_multi}

\begin{table}[t]
  \centering
  \begin{tabular}{ccll}
    \toprule
   chain id & $m^f_0$ & $\kappa$ & $am_\mathrm{q}$ \\
    \midrule
    D2 & $m^0$ & 0.1352 & 0.016 \\
       & $m^1$ & 0.128  & 0.224 \\
       & $m^2$ & 0.115  & 0.665 \\
    \bottomrule
  \end{tabular}
  \caption{The parameters used in the FS stochastic estimator for the
    light-quark loop. The critical hopping parameter is taken from
    ref.~\cite{Luscher:1996ug}.}
  \label{tab:ym_masses}
\end{table}
\begin{figure}[t]
  \centering
  \includegraphics[scale=0.8]{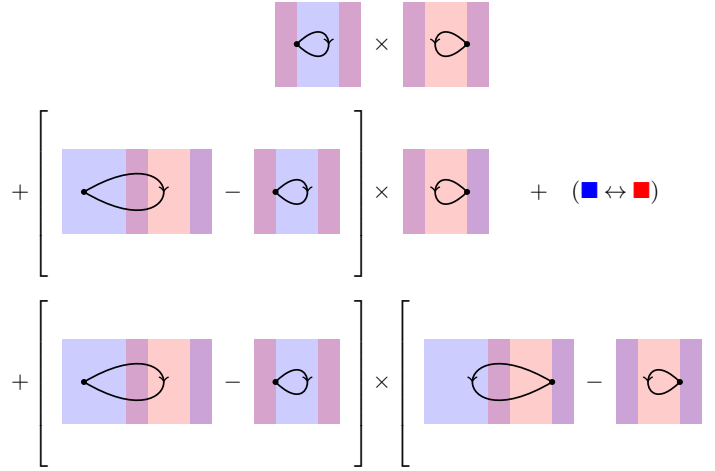}
  \caption{Graphical illustration of the unbiased estimator for the
    disconnected contribution to the meson two-point function in a two-level
    simulation.
    The first line is fully factorized,while the following two depend on the
    whole gauge field and require independent measurements for each combination
    of level-1 configurations for a multi-level estimator.}
  \label{fig:disc_fact}
\end{figure}
In this section, we apply the FS estimator proposed in the previous subsection
to the disconnected two-point function in the two-level simulation described
in subsection~\ref{sub:vector_ml}.
We test an unbiased multi-level  estimator for the disconnected
diagram~\cite{Ce:2016idq},
\begin{align}
  C_{\Gamma_d}(y_0,x_0) &=
      \frac{a^6}{L^3}\sum_{\bm{y},\bm{x}}\eval{
          \tr{\gamma_5\Gamma Q^{-1}(y,y)}\tr{\gamma_5\Gamma Q^{-1}(x,x)}}
  \label{eq:disc_twopt1}
\end{align}
where $x\in\Omega_0^*$, $y\in\Omega_1^*$, which is constructed by
decomposing the hermitian propagator, $Q^{-1}$, via the identity
\begin{align}
  \tr{\gamma_5\Gamma Q^{-1}(x,x)} &=
    \tr{\gamma_5\Gamma Q^{-1}_{\Omega^*_i}(x,x)}
    + \tr{\gamma_5\Gamma(Q^{-1} - Q^{-1}_{\Omega^*_i})(x,x)},
  \label{eq:disc_ml}
\end{align}
in each of the blocks $x\in\Omega_i^*$ for $i=0,1$.
This decomposition results in three contributions to the disconnected diagram,
\begin{align}
  C_{\Gamma_d}(y_0,x_0)
          &= \frac{a^6}{L^3}\sum_{\bm{y},\bm{x}}\eval{(
            W_{\Gamma_d}^{(\mathrm{f})}
            +W_{\Gamma_d}^{(\mathrm{r}_1)} 
        +W_{\Gamma_d}^{(\mathrm{r}_2)})(y,x)},
  \label{eq:disc_twopt2}
\end{align}
which are illustrated graphically in the first, second and third lines of
fig.~\ref{fig:disc_fact}.
The propagators $Q^{-1}_{\Omega_i^*}$ are represented by
the the small blocks, while the the propagators on the whole
lattice, $Q^{-1}$, are represented by the large blocks.

The contribution on the first line of the figure is therefore fully
factorized, as the propagators depend only on the gauge field in the level-1 
configurations respectively, and requires only $n_0\times~n_1$ measurements
for a multi-level estimator.
The other contributions depend on the gauge field in the whole lattice, and if
no further factorization is introduced, a multi-level estimate would require
measurements on each of the $n_0\times n_1^2$ combinations of level-1
configurations.
Alternatively, if the variance on these non-factorized pieces is suppressed,
then it may be sufficient to estimate these corrections using fewer, e.g.
$n_0\times n_1$, measurements and the factorization strategy within a
multi-level scheme will still be effective.

\begin{figure}[t]
    \centering
    {\includegraphics{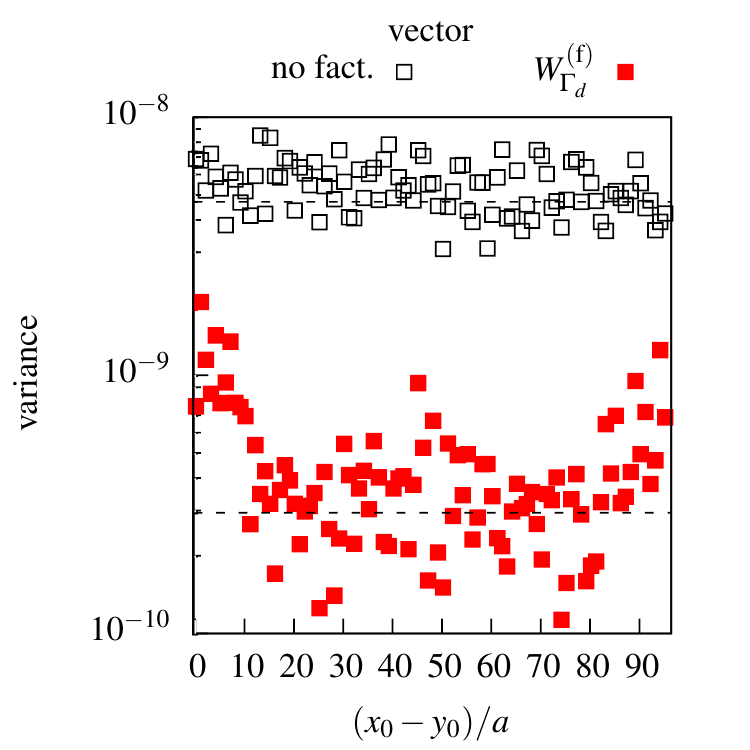}}%
    {\includegraphics{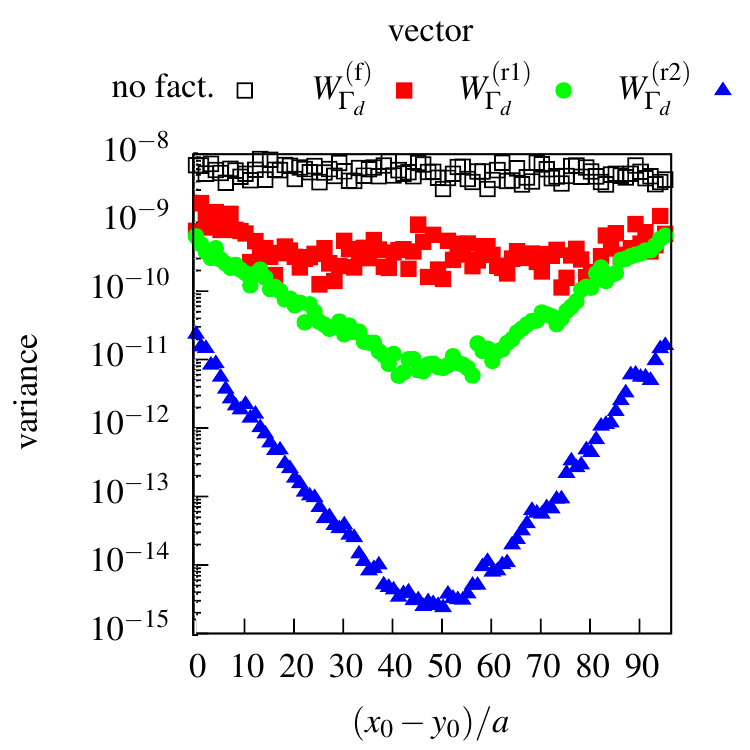}}
    \caption{The variance of the vector channel disconnected diagram as a
    function of the source-sink separation. In the left-hand panel the
  variance of the non-factorized estimator using $n_0\times n_1$
  configurations is compared with the factorized contribution with two-level
  integration (red filled squares), showing the expection $1/n_1=1/16$ reduction
  in the variance (dashed lines).
  In the right-hand panel, the variances of the remaining contributions (green
  and blue points) computed with $n_0\times n_1$ configurations is shown to be
suppressed with respect to the factorized contribution.}
    \label{fig:ml_var}
\end{figure}
Preliminary results for the variance with and without two-level integration
are shown for the vector channel in fig.~\ref{fig:ml_var}.
In the left-hand panel, the variance of the disconnected contribution as a
function of the separation $x_0-y_0$ without factorization using $n_0\times
n_1$ measurements (black open squares) is compared with the variance on the
factorized contribution using two-level integration (red filled squares).
As expected the variance is constant in time, which gives rise to an
exponentially bad signal-to-noise problem.
As depicted by the dashed lines, the variance on the factorized contribution
is reduced by $n_1=16$ using the two-level simulation.
In the right-hand panel, the variance on the two remaining contributions are
shown with the green and blue points.
The variance is sufficiently suppressed on these remaining
contributions that they remain subleading.
Furthermore, the variances on these contributions appears to be exponentially
suppressed with the distance.

These preliminary results indicate that a two-level scheme for the
disconnected diagram in the vector channel can be easily implemented and the
dominant uncertainity arises from a fully factorizable contribution.

\section{Conclusions}
\label{sec:conclusions}

In these proceedings we have proposed new estimators for both connected and
disconnected contributions to meson two-point functions suitable for
multi-level integration, and tested them in a two-level quenched 
simulation.
Such estimators could be applied to tackle the signal-to-noise problem in, for
example, singlet pseudoscalar spectroscopy, or the leading-order
hadronic vacuum polarization.

For the connected diagram, a stochastic estimator for the factorized
contribution is proposed which is computed sequentially in each region of the
two-level simulation.
As expected, the factorized contribution to the connected two-point function
is a good approximation to it, and furthermore the error on the remainder is
small.
In the two-level simulation we have observed the expected gain over the
single-level simulation when the sink is adequately far from the frozen
region.
The dependence of the gain on the distance from the frozen region is currently
under investigation.

In the disconnected sector we have proposed a new stochastic estimator for the
quark loop which is applicable with or without multi-level integration.
This estimator is similar in spirit to the frequency-splitting of the quark
determinant performed in the HMC, where the separation allows the dominant
fluctuations to be controlled because they are less computationally intensive.
In addition, we have introduced an alternative estimator for the difference of
quark loops with different quark masses, which provides a signficant
improvement, and could be applied to the disconnected contribution to the HVP.
A detailed theoretical description of the variances using the estimators can
be found in ref.~\cite{Giusti:2019inprep}.
Like the connected diagram, we find the variance of the factorized
contribution to be dominant, which means that a factorized multi-level
strategy is applicable.

The results of these new techniques confirm those presented in
the previous Lattice conference~\cite{Ce:2017ndt}, which were obtained with
open boundary conditions, and extends them to the disconnected diagram in the
vector channel.
We stress that the magnitude of the correction depends heavily of the quark
mass, and therefore the application of these techniques to QCD with light
quarks is not straightforward.

\paragraph{Acknowledgements} 
Simulations have been performed the PC-cluster Marconi at CINECA (CINECA-INFN
and CINECA-Bicocca agreements), and on the 
PC-cluster Wilson at Milano-Bicocca. We thankfully acknowledge the computer
resources and technical support provided by these institutions.

\printbibliography
\end{document}